\documentclass[12pt,preprint]{aastex63}

\usepackage{natbib}
\setcitestyle{super} 

\newcommand{\xco}{\,\hbox{$X_{\rm CO}$}}
\newcommand{\Nco}{\,\hbox{$N_{\rm CO}$}}
\newcommand{\Nhtwo}{\,\hbox{$N_{\rm H_2}$}}
\newcommand{\Ntot}{\,\hbox{$N_{\rm tot}$}}
\newcommand{\NhtwoLOS}{\,\hbox{$N_{\rm H_2}^{\rm LOS}$}}
\newcommand{\nhtwo}{\,\hbox{$n_{\rm H_2}$}}
\newcommand{\nh}{\,\hbox{$n_{\rm H}$}}

\newcommand{\Tex}{\,\hbox{$T_{\rm ex}$}}
\newcommand{\Tkin}{\,\hbox{$T_{\rm kin}$}}

\newcommand{\msun}{\,\hbox{$M_{\odot}$}}

\newcommand{\kms}{\,\hbox{\hbox{km}\,\hbox{s}$^{-1}$}}

\newcommand{\htwo}{\,\hbox{$\rm{H_ 2}$}}

\newcommand{\ang}{\,\hbox{\AA}}

\newcommand{\feii}{\,\hbox{[\ion{Fe}{2}}]}
\newcommand{\sii}{\,\hbox{[\ion{S}{2}}]}
\newcommand{\nii}{\,\hbox{[\ion{N}{2}]}}
\newcommand{\oiii}{\,\hbox{[\ion{O}{3}]}}

\newcommand{\cooz}{\hbox{$\rm CO(1$-$0)$}}
\newcommand{\coto}{\hbox{$\rm CO(2$-$1)$}}

\newcommand{\cott}{\hbox{$\rm CO(3$-$2)$}}
\newcommand{\coft}{\hbox{$\rm CO(4$-$3)$}}

\newcommand{\hcopoz}{\hbox{$\rm HCO^+(1$-$0)$}}
\newcommand{\hcopft}{\hbox{$\rm HCO^+(4$-$3)$}}

\newcommand{\bff}{\,\hbox{$b_{ff}$}}

\shorttitle{}
\shortauthors{}

\begin{document}

\title{\Large Linking pressure gradients with the stability of molecular clouds in galactic outflows}
\maketitle

{\vskip -6pt
\noindent
\bf K. M. Dasyra$^{1,2}$, G. F. Paraschos$^{3,2,1}$, T. G. Bisbas$^{4,2}$, F. Combes$^{5}$, J. A. Fern\'andez-Ontiveros$^{6}$

\small \rm \flushleft $^1$ Section of Astrophysics, Astronomy \& Mechanics, Department of Physics, National and Kapodistrian University of Athens, Panepistimioupolis Zografou, 15784 Athens, Greece\\
$^2$ Institute for Astronomy, Astrophysics, Space Applications, and Remote Sensing, National Observatory of Athens, 15236 Penteli, Greece\\
$^3$ Max-Planck-Institut f\"ur Radioastronomie, Auf dem H\"ugel 69, 53121, Bonn, Germany\\
$^4$ Physikalisches Institut, Universit\"at zu Köln, Z\"ulpicher Stra{\ss}e 77, 50937 K\"oln, Germany\\
$^5$ Observatoire de Paris, LERMA, Coll\`ege de France, PSL, CNRS, Sorbonne University, Paris, France\\
$^6$ Istituto di Astrofisica e Planetologia Spaziali, Via Fosso del Cavaliere 100, I--00133, Roma, Italy

\vskip 12pt}

{ \noindent \bf 

The jets launched by actively accreting black holes are capable of launching several of the massive (million or billion solar mass) molecular outflows observed in galaxies. These outflows could suppress or enhance star formation in galaxies. To investigate the stability of clouds capable to form stars in outflows, we modeled CO and HCO$^+$ ALMA data of the galaxy IC5063, in which black-hole jets impact molecular clouds. Using a radiative transfer code that self-consistently performs astrochemical and thermal balance calculations based on the available gas heating sources, we found that mechanical heating and cosmic ray (CR) heating are fully capable of individually reproducing the data. In our best-fit model, CRs provide $\sim$1/3rd of the dense gas heating at the radio lobes, emphasizing the role of this often neglected mechanism in heating the gas and potentially generating outflows. The gas temperature and density indicate that the jet passage leads to an increase of $\sim$1 order of magnitude in the internal pressure $P_i$ of molecular clouds (with $P_i$/$k$ from 8$\times$10$^5$ up to 7$\times$10$^6$ K cm$^{-3}$), irrespective of the excitation mechanism. From the fluxes of \sii\ and \nii\ lines in VLT MUSE data, the external pressure $P_e$ of molecular clouds increases in several regions enough to exceed P$_i$. This result 
leads us to conclude that we are observing the expansion of an ionized overpressurized cocoon that compresses molecular clouds and that could lead to their collapse. Some jet-impacted clouds, nonetheless, near pathways that the jet cleared have increased P$_i$ and decreased P$_e$. They are likely to undergo evaporation of their outer layers. Part of the evaporated layers could mass load the outflow thanks to ram pressure from co-spatial ionized gas flows. The observed pressure changes thus suggest that both star formation enhancement and suppression could simultaneously occur.
\\
}

In the last decade, numerous detections of outflows, i.e., winds of molecular gas, revealed that a high fraction (often 1-20\%) of the gas content of dense clouds in galaxies can be kinematically and dynamically perturbed \citep{fluetsch19,veilleux20}.
These detections raised questions about the suppression \citep{bolatto13,dasyra16,aalto16} or enhancement  \citep{croft06,crockett12,maiolino17} of star formation in galactic regions with outflows due to the rarefaction or compression of clouds. An excellent laboratory to investigate for hydrodynamical changes of clouds in outflows is the galaxy IC5063. The collision of  radio-emitting jets with dense interstellar medium (ISM) clouds near the nucleus of IC5063 results in the initiation of several discrete, spatially offset outflows along or near the jet trail. These outflows have been seen in \feii , \htwo , CO and other lines \citep{dasyra15}. Even the expansion of a jet-inflated cavity (a.k.a. cocoon) has been proposed based on HI \citep{oosterloo00} and CO \citep{morganti17} data. 
A first study of the molecular gas excitation based on the \coft /\coto\ flux ratio revealed a significant temperature increase from ambient (not-jet-impacted) to jet-impacted regions. Some molecular gas in the outflow has temperatures $>$100K
 \citep{dasyra16}. It is, thus, plausible that, along with the change in the excitational state of jet-impacted clouds, a change in their dynamical state also occurred. \\

The stability, expansion, or contraction of clouds is determined by the interplay of their self gravity with changes in their internal and external pressure. Despite the importance of pressure in this process, pressure maps have not been presented before for the molecular gas and its surrounding medium in regions with outflows. They have only been presented for the X-ray emitting gas,  delineating the surfaces of jet-inflated bubbles, such as the cavities in Perseus A \citep{fabian06} or the Fermi bubbles in the Milky Way \citep{ponti19}. Here, we present pressure maps of the tenuous ionized medium surrounding molecular clouds in the nucleus of IC5063 from VLT MUSE and SINFONI data. We also present pressure maps of the molecular gas itself, computed from the deepest available CO and HCO$^+$ ALMA cubes, which we reconstructed from all available data in the ALMA archive. Rotational lines of CO from J=1-0 up to J=4-3 and HCO$^+$ J=1-0 and J=4-3 were fitted for the gas density, temperature, and pressure derivation in a spatially-resolved, pixel-by-pixel manner. The resulting pressure maps depend little on the mechanism of energy deposition into the dense interstellar medium (ISM), whether it be mechanical heating or cosmic-ray (CR) heating or both. \\

Mechanical heating is a widely examined mechanism that encapsulates a series of physical processes: radiation pressure drives winds of tenuous ionized gas, which then interact with molecular clouds via ram pressure, hydrodynamic and thermal instabilities that assist ram pressure by cloud evaporation, and cloud uplifting done by the work of adiabatically expanding plasma bubbles  \citep{hopkins_elvis10,zubovas14, dannen20}. These mechanisms also apply to any wind initiated by a radio jet \citep{wagner16}. Another mechanism present in any wind, and in particular in a jet-driven wind, are cosmic rays (CRs). CRs are less frequently discussed in the literature as a mechanism of gas heating, acceleration and wind launching \citep{ruszkowski17}, even though they can be energetically important: in the solar neighbourhood, their energy density is comparable to the thermal energy density of the ISM \citep{klessen_glover}. CRs penetrate clouds, directly interacting with the dense gas at great cloud depths via ionizing collisions that can even change the gas chemistry \citep{papadopoulos10,padovani18,gaches19}.
Hydrodynamical simulations indicate that CR-driven winds around supernovae (SN) are cooler and denser than thermally-driven winds \citep{girichidis16}. The analysis that follows shows that both mechanical and CR heating are capable of reproducing the observed CO and HCO$^+$ spectral line energy distribution (SLED) in IC5063. Both mechanisms lead to emission from excited cloud layers with such kinetic temperatures,  \Tkin , and \htwo\ volume densities, \nhtwo , that correspond to similar pressures.  \\

To model SLEDs, we ran radiative transfer (RT) models that were carried out in a self-consistent manner for the available gas heating sources and that only considered dynamically meaningful cloud states with respect to their volume density, column density, and kinetic temperature combination (see the methods for more details).  
The code we used was 3D-PDR \citep{bisbas12}, which performs thermal balance calculations for the input heating sources and solves the astrochemical network equations prior to carrying out radiative transfer and outputting line fluxes. We used two heating sources of galactic origin in all our models: stellar UV photons and stellar-evolution-related CRs. CRs are included in 3D-PDR via the ISM ionization rate that they cause, $\zeta_{CR}$. We also used UV photons related to the AGN and two heating sources related to the jet: more CRs and mechanical heating. The additional CR flux is a highly attractive yet little explored candidate for the dense gas acceleration and excitation in radio galaxies.
The mechanical heating is associated with a widely implemented feedback mode in hydrodynamical simulations of cosmological scales \citep{dubois12} or galaxy scales \citep{gaibler12,wagner12}. Indeed, a contribution of all these mechanisms to the ISM heating rate, $\Gamma$, was indicated by our best-fit model.\\

\begin{figure*}[!t]
	\begin{center}
		\includegraphics[width=16.5cm]{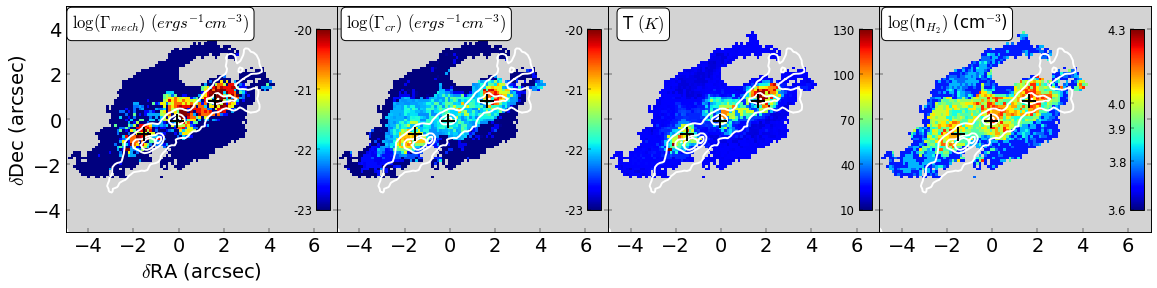}
	\end{center}
	\vspace{-0.6cm}
		\caption{ Spatially-resolved CO and HCO$^+$ SLED fitting results: Mechanical heating rate, cosmic ray heating rate, kinetic temperature, and \htwo\ volume density maps. Pixels with S/N ratio $<$3 in CO (1$-$0), (2$-$1) or (3$-$2) were masked out in this and all following relevant figures. Contours mark the jet trail, as traced mainly by the \oiii\ emission in a {\it Hubble Space Telescope} Wide Field Planetary Camera 2 narrow-band image. Crosses mark the position of the radio core and lobes from the ALMA continuum. 
		}
		\label{fig:heating}
\end{figure*}

In our best-fit model, UV radiation is mainly important for ambient regions, far from the radio lobes (e.g., in the northern spiral arm of Fig.~\ref{fig:heating}). There, UV heating rates of a few 10$^{-22}$ erg s$^{-1}$cm$^{-3}$ are achieved thanks to unattenuated UV fluxes $f_{UV}$ that are 5-10 times greater than those in the solar neighbourhood \citep{draine78}. 
The heating rate due to CRs, $\Gamma_{CR}$, which can mainly be attributed to CRs of galactic origin, is there only at a level of $\sim$10$^{-24}$ erg s$^{-1}$cm$^{-3}$, despite a typical ionization rate similar to that of the solar neighbourhood ($\zeta_{CR}$ $\sim$5$\times$10$^{-17}$ s$^{-1}$). However, the role of CRs becomes important at the radio lobes, where the most intense jet-dense ISM interaction takes place (Fig.~\ref{fig:heating}). At the north-west lobe, $\Gamma_{CR}$ increases to 6$\times$10$^{-21}$ erg s$^{-1}$cm$^{-3}$ thanks to $\zeta_{CR}$ reaching 2$\times$10$^{-14}$ s$^{-1}$. Still, mechanical heating is the dominant mechanism throughout the entire radio jet trail, with $\Gamma_{m}$ reaching values as high as 2$\times$10$^{-20}$ erg s$^{-1}$cm$^{-3}$. Within one beam from the north-west radio lobe, mechanical heating contributes $\sim$3 times more than CRs to the heating budget. Taking into account all mechanisms, $\Gamma$ overall increases by a factor of 200 from ambient regions to the jet vicinity.
 The extra heating results in an increase of the jet-impacted gas temperature: \Tkin\ increases from 20\,K in ambient regions to 130\,K at the north-west lobe (Fig.~\ref{fig:heating}), with a corresponding increase of the \cooz\ excitation temperature, \Tex, from 10\,K to 110\,K. The heated molecular gas is also denser (with \nhtwo $\sim$20000 cm$^{-3}$ instead of 3000 cm$^{-3}$), either because it was denser prior to the jet passage, or because its deeper and denser layers got heated or exposed after the jet passage.  The detection of \hcopft\ in the same region \citep{oosterloo17}, which requires densities $\gtrsim$10$^{7}$ cm$^{-3}$ for collisional excitation, suggests that the jet heats the clouds all the way down to their cores. The line-of-sight-integrated, intrinsic
 \htwo\ column density of clouds \NhtwoLOS\ there exceeds 10$^{23}$ cm$^{-2}$. It is above the threshold for collapse and star formation \citep{kennicutt98}, and two orders of magnitude greater than that of the ambient ISM. Even the densest gas phase that could form stars, gets, thus, heated by the jet.  \\

While our best fit model points to a joint contribution of CR and mechanical heating to the energy budget, either mechanism can reproduce the CO and HCO$^{+}$ SLEDs of jet-impacted regions alone, as demonstrated by two more models shown in the methods: one with $\zeta_{CR}$=free and $\Gamma_{m}$=0 and another with $\Gamma_{m}$=free and Galactic $\zeta_{CR}$=10$^{-16}$\,s$^{-1}$. Each mechanism individually leads to a similar increase in $\Gamma$, for a reasonable increase of $\Gamma_m$ (in agreement with the outflow kinetic luminosity) or of $\zeta_{CR}$ (in agreement with values in the Galactic center or in the shock fronts near SN remnants \citep{schuppan12}). This result underlines the importance of incorporating CRs in outflow studies. Our conclusions could concern all outflows: jet-driven, AGN-driven and starburst-driven, as CRs can be found in all of them because of Fermi acceleration. 
Given that CR heating is likely associated with momentum exchange, CRs could help the launching of outflows, potentially diminishing the role of the adiabatic expansion of radiation-heated plasma bubbles to momentum boosting. Even in cosmological simulations with jet-related feedback, the inclusion of CRs along with mechanical heating could be essential for the proper derivation of galaxy masses.  \\

Differences between the models also exist. The CO emission of CR-heated gas originates from cooler and denser regions, with \Tex $\sim$75-80K, \nhtwo $\sim$10000-16000\,cm$^{-3}$, than that of mechanically-heated gas, with \Tex $\sim$90-130K, \nhtwo $\sim$7000-10000\,cm$^{-3}$. Indicative errors on these quantities are of order 20-30\%. The conversion factor of the CO luminosity at the surface of clouds to the \htwo-mass in their interior, \xco , differs too. For mechanical heating, \xco\ drops from nearly Galactic values at ambient regions to $\sim$4-5 times lower values along the jet trail, where dispersed accelerated gas exists \citep{bolatto13,dasyra16,richings18}. For CR heating, \xco\ is nearly Galactic everywhere, as the emission from dense layers adds to the emission from dispersed layers. This result is of extreme importance for the mass measurement of jet-impacted clouds. A promising diagnostic of the most appropriate model is the flux ratio of CO(7$-$6) to CI(2$-$1), i.e., of two lines that can be simultaneously observed near 809 GHz. In several regions along the jet trail, CO(7$-$6) is brighter than CI(2$-$1) for mechanical heating, but CI(2$-$1) is brighter than CO(7$-$6) for CR heating (see the methods). This inversion owes to the ionization of He by CRs. He$^+$ then engages in the reaction He$^+$+CO$\rightarrow $O+C$^+$+He, efficiently destroying CO molecules and enhancing atomic C. Finally, C$^+$ easily captures an electron, becoming neutral \citep{bisbas15}. It is noteworthy that, in general, X-rays could be leading to similar chemistry an 
contributing to $\Gamma$ \citep{meijerink05}. In IC5063, diffuse X-ray emission linked both to the weak AGN and to the jets was recently discovered \citep{travascio21}, but it is unlikely to play a dominant role on the gas excitation. A gas heating rate of 10$^{-20}$ erg s$^{-1}$ cm$^{-3}$ could only be attained at the north-west radio lobe if all clouds are near (e.g., within several parsecs away from) a single X-ray source of integrated luminosity $\sim$ 10$^{39}$ erg s$^{-1}$ \citep{maloney96}. A more realistic spatial distribution of the molecular gas, e.g., including clouds hundreds of pc away from the X-ray emitters, would considerably decrease $\Gamma_{X rays}$, demoting or eliminating the role of X-ray heating. Still, we note that even if X-rays were fully capable of providing the needed heating rate, they would lead to similar SED as CRs \citep{meijerink05}. For these reasons, we do not implement them further but instead focus on the role of the abundant jet-related CRs. \\

\begin{figure*}[!t]
	\begin{center}
	    \includegraphics[width=17.cm]{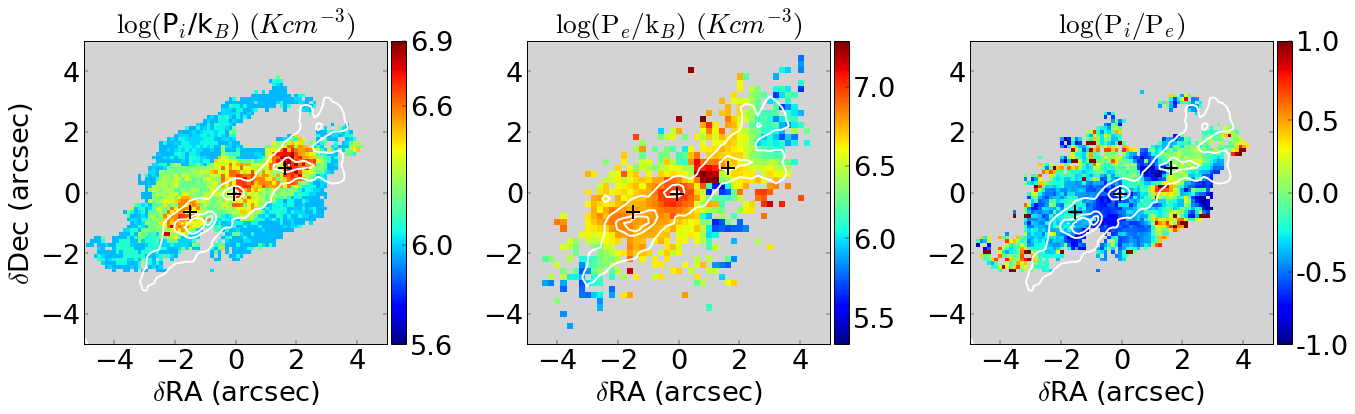}
	    \end{center}
    \vspace{-0.5 cm} 
        \caption{ {\it Left:}
         Internal pressure of typical line-of-sight molecular clouds,  deduced from CO and HCO$^+$. {\it Middle}: Ionized medium pressure, considered external to molecular clouds, from \sii\ and \nii .
         {\it Right:} Ratio of the two former pressures, showing that due to the jet, P$_e$ can be either higher or lower than P$_i$. 
          }
	\label{fig:pressure}
\end{figure*}

The internal pressure P$_i$ of a typical line of sight cloud is nearly identical for all excitation models. This is a very powerful result for studies of cloud stability, which owes to lower \Tkin\ but higher \nhtwo\ values when the gas is CR heated instead of mechanically heated. In three dimensions, P$_i$ is 5\nhtwo $k$\Tkin/2\,+\,3\nhtwo m${_H}$U$_{turb}^2$, where $k$ is the Boltzmann constant and U$_{turb}$ is the cloud turbulent velocity, 
0.6$\kms$. P$_i$ displays significant spatial variations along the jet trail (Fig.~\ref{fig:pressure}). Unaffected gas clouds at ambient locations have P$_i$/$k$=8$\times$10$^5$ K cm$^{-3}$. Heavily impacted clouds at the radio lobes have P$_i$/$k$=7$\times$10$^6$ K cm$^{-3}$, which is similar to the values of giant molecular clouds in the distant Universe \citep{dessauges19}, and which indicates different conditions for future star formation. Under the assumption that all clouds had similar initial properties, the spatial gradients indicate how the internal pressure of individual clouds changed after the jet passage. Unaffected clouds kept memory of the initial properties, while impacted clouds significantly changed properties. Indeed, the ratio N$_{tot}$/\nhtwo ,  which is a measure of radius, is two times larger at ambient regions (0.8\,pc) than at the north-west radio lobe (0.4\,pc). \\

 Along with internal pressure gradients, external pressure gradients will determine the hydrodynamic fate of molecular clouds.  The external medium pressure P$_e$ is assumed to be that of the ionized gas seen in VLT MUSE data. We used classical optical line diagnostics, \sii\,6716\ang /\sii\,6732\ang\ and (\nii\, 6548\ang +\nii\,6584\ang)\,/\,\nii\,5758\ang\ \citep{luridiana15}, to determine the ionized gas density and temperature, respectively, under the assumption of thermal equilibrium for the electronic level populations of atoms. We computed P$_e$ as (3/2)n$_H k$T and found that P$_e$/$k$ can reach 2$\times$10$^7$ K cm$^{-3}$, peaking near an ionized gas bifurcation point seen in \oiii\ \citep{dasyra15}. It indicates a severe jet-cloud collision there. 
 The drop happens along the axis of a known biconical \feii\ and \htwo\ outflow \citep{dasyra15}. Coaxial to this outflow, there are regions with low ionized gas density (see the methods) that can be explained as pathways that the jet opened through the ISM once scattered at the impact point. Indeed, faint 3\,cm emission from the jet is seen  parallel to these regions, at a direction perpendicular to its initial path \citep{oosterloo00}. With the exception of these regions, the ionized and molecular media have otherwise surprisingly comparable pressures, with P$_e$ being on average two times P$_i$. Pressure measurements of the gas  from VLT SINFONI data yielded a similar range of values at the radio lobe vicinity. From \feii\ 1.533 and 1.644 $\micron$, we deduced shocked ionized medium densities of 300-1100 cm$^{-3}$ \citep{pradhan_zhang93, luridiana15}. From  \htwo\ (0-0) S(1) and S(3), we obtained  \Tkin\ of 1500-2000K for the thermalized molecular gas, noting that this  range could be somewhat elevated if there is a significant contribution of fluorescence to the upper quantum level populations of many molecules. Combining these numbers, even though they reflect properties of different ISM phases that may not coincide or be in equilibrium, yields P$_{shocked}$/$k$ $\sim$10$^6$-10$^7$ K cm$^{-3}$. \\

With P$_e$ frequently exceeding P$_i$ (Fig.~\ref{fig:pressure}), we are observing a phase of expansion of an overpressurized ionized medium bubble, the so called cocoon of the jet, which engulfs and compresses clouds as it propagates in all directions. Otherwise, P$_i$ would exceed P$_e$, and the stability of molecular clouds would be achieved thanks to self gravity, as in the normal ISM. Whether the compression leads to a temporary enhancement of star formation is unknown, depending on other factors such as microturbulence. The overpressurized region would be more extended if the microturbulence of the ionized gas exceeded that of the molecular gas. Simultaneously, underpressurized regions exist along the pathways that the jet opened up. In clouds with kinetic energy exceeding the virial value, which happens in most regions with P$_i$/P$_e$ $\gtrsim$2 (Figure~\ref{fig:pressure}), evaporation of cloud layers that become gravitationally unbound is possible. For the region $\sim$350\,pc west of the nucleus, the evaporation will be possible even the ionized gas microturbulence is 10\kms . It could be followed by mass loading of the outflow with the aid of ionized winds. Direct proof that cloud evaporation has occurred is the decrease of the optical depth along the jet trail \citep[][methods]{dasyra16}.  Conditions of both suppression and enhancement of star formation are, thus, simultaneously plausible.

\clearpage

\section{Methods}

\subsection{The Atacama Large Millimeter Array CO emission maps}
\label{sec:data}

To create two-dimensional maps of the CO spectral line energy distribution (SLED), we used all archived Atacama Large Millimeter Array (ALMA) data of IC5063. For the \cooz\ and \cott\ cube reconstruction, we used the observations for the programs 2015.1.00420.S (PI Combes) and 2015.1.00467.S (PI Morganti) with total on-source integration times of 2.5\,hrs. For \coto , we merged the data of the programs 2012.1.00435.S (PI Morganti) and 2016.1.01279.S (PI Carniani) with a total on-source integration time of 1.5\,hrs. For the second program, we used the central pointing of the mapping observations. For \coft , only data from the 2015.1.00420.S (PI Combes) program were available, with an on-source integration time of 40\,mins. 

The reduction was carried out with the Common Astronomy Software Applications (CASA). To calibrate the data of the individual programs, we used the respective CASA version and functions that were used by the ALMA staff. For all following tasks, we used CASA version 5.4.0-70.  We ran the function ``{\tt uvcontsub}" for each measurement set, employing zero-order polynomials to model and subtract the continuum emission at frequencies free of line emission. For programs with spectral coverage overlap, we used the functions ``{\tt mstransform}" and ``{\tt msconcat}" to merge the measurement sets, once we regrided them to the common velocity grid and resolution. For the \coto\ data, we also ran the ``{\tt statwt}" function to recompute the visibility weights of the 2012 data before concatenating them with the 2016 data, as needed for
CASA versions prior to 4.2.2.

To  transform the merged visibilities into images we used the ``{\tt tclean}" function, running it iteratively to clean the images from the synthesized beam sidelobes. The primary beam response correction was applied. To ensure that the extended flux is well recovered, in particular for the low-frequency lines, we opted to use a Briggs robustness parameter of 2, corresponding to natural weighing of the uv plane. This choice mattered 
for the \cooz\ flux, but made no difference for the measured \coft\ flux. The image reconstruction with natural weighing led to synthesized beams of 0.72$\arcsec$$\times$0.65$\arcsec$, 0.63$\arcsec$$\times$0.59$\arcsec$, 0.66$\arcsec$$\times$0.56$\arcsec$, and 0.36$\arcsec$$\times$0.33$\arcsec$  for CO (1$-$0), (2$-$1), (3$-$2), and (4$-$3), respectively. To reach a common resolution for all lines, we convolved the cubes to the \cooz\ beam using the function ``{\tt ia.convolve2d}". For the adopted cosmology ($\Lambda$CDM, with H$_0$=70 \kms\  Mpc$^{-1}$, $\Omega_{M}$=0.3 and $\Omega_{\Lambda}$=0.7) and for the measured source redshift (0.01128), the beam corresponds to a scale of 165$\times$150\,pc$^2$.
All cubes were built 
at a common spectral resolution channel of 20\kms .  In the common spatial and spectral resolution, the sensitivity levels were 0.25, 0.23, 0.39, and 1.8 mJy/beam for CO (1$-$0), (2$-$1), (3$-$2), and (4$-$3), respectively. For the inner 5 kpc (22\arcsec ), the respective line fluxes were 30.6($\pm$2.4) Jy \kms , 77.0($\pm$4.5) Jy \kms , 107($\pm$8.0) Jy \kms ,  and 140($\pm$20) Jy \kms . 

Observations of \hcopoz\ and \hcopft , with respective exposure times of 38\,mins and 82\,mins, were acquired for the same programs and were used to constrain the CO SLED excitation modeling. The HCO$^+$ data reduction was carried out similarly to that of the CO. The emission was more compact than \coft\, allowing us to adopt a robustness parameter of 1 for the image reconstruction. We then convolved the cubes to the resolution of \cooz . The measured line fluxes were 5.6($\pm$0.8) Jy \kms\ for \hcopoz , 3.2($\pm$0.2) Jy \kms\ for \hcopft . The noise, at the common resolution, was 0.4 and 0.5 mJy/beam, respectively. Information on the HCO$^+$ line extent and ratio is given in Fig.~\ref{fig:line_input}.

We created moment maps of the cubes using the weighted-flux algorithm that is described in the CASA ``{\tt immoments}" function, with different integration limits around each spectral line per pixel. While ``{\tt immoments}" uses a fixed velocity range for all pixels, our routine determines the integration limits in each pixel by performing an initial Gaussian line profile fit. 
This way, the resulting moment maps suffer less from noise contamination in regions with low dispersion (e.g., in the spiral arms) and include high-velocity gas emission in regions with outflows. The derived line fluxes, recession velocity and velocity dispersion maps, averaged over all CO lines, are shown in  Fig.~\ref{fig:line_input}. All maps include both the disk and outflow emission, as a global solution is sought: the emission of the quiescent gas in the disk cannot be  distinguished from that of the jet-heated or jet-accelerated gas perpendicularly to the line of sight. Prior to fitting the data, we divided the observed line fluxes by their full width half maximum in each pixel, so that all computations are performed for individual molecular clouds of intrinsic line width of 1\kms\ in the line of sight.

\subsection{Radiative transfer modeling of the CO SLED}
\label{sec:model}

To model any SLED, RT for spectral lines is needed. The most traditional and simple RT approach is the modeling of one molecular species at a time, using fixed abundance ratios between species, and assuming that the gas is isothermal (as in, e.g., RADEX). Modeling of molecular line fluxes with RADEX \citep{vandertak07} has been presented before for IC5063 \citep{oosterloo17}, indicating that the plausible range of \Tkin\ and \nhtwo\ increases with proximity to the radio lobes. However, not all pre-selected gas column and volume density combinations correspond to dynamically meaningful cloud states (e.g., some solutions correspond to collapsed or evaporated states). Likewise, pre-selected temperatures and abundances do not necessarily match their appropriate values for the available gas heating sources (and heating rate). Moreover, the isothermal assumption is unrealistic. \\

To improve the RT handling we opted to use a code that implements heating and cooling calculations until thermal balance is self-consistently reached, together with astrochemical calculations that provide the appropriate abundance ratios for the enhancement or destruction of species at each temperature. For this purpose, we used the code {\sc 3d-pdr} \citep{bisbas12}, which is publicly available at https://uclchem.github.io/3dpdr/. The astrochemical calculations used for the modelling of the observed CO and HCO$^+$ line intensities depend on the implemented heating sources. We included far-ultraviolet radiation heating, CR heating, and mechanical (shock-related) heating. Considering various cooling processes, {\sc 3d-pdr} converges once the total heating and cooling functions are balanced. It uses a subset of the UMIST2012 chemical network \citep{mcelroy13} comprising 33 species and 330 reactions. The metallicity is assumed solar, the dust-to-gas ratio is 100,
and the initial gas-phase chemical abundances, normalized to hydrogen, are $\rm [C^{+}/H]=1\times10^{-4}$ \citep{cardelli96}, $\rm [O/H]=3\times10^{-4}$ \citep{cartledge04} and $\rm [He/H]=0.1$. The thermochemical grid is built upon one-dimensional uniform-density distributions. Taking the level populations provided by the astrochemical calculations, the code performs RT calculations to compute the coolants' emission. It is assumed that the cloud is embedded in the cosmic microwave background radiation field and that the molecules get excited by collisions with up to seven partners (e.g., hydrogen, helium). The collision rates are taken from the LAMBDA database \citep{schoier05}. The calculations are performed under the large velocity gradient (LVG) approximation for an expanding sphere. The code provides the kinetic temperature {\Tkin}, the excitation temperature \Tex , the radiation temperature $T_R$, the abundance and the volume density of each requested coolant.  The coolants that we used were CO and HCO$^+$. Simultaneous solutions of the CO SLED and HCO$^+$ (4-3)/(1-0) ratio were found. \smallskip

To construct the grid of viable models, we set the input parameters that relate to the gas excitation as follows. We used GALEX near and far ultraviolet flux measurements,  1.2$\times10^{-4}$ and 5.2$\times10^{-5}$\, Jy, respectively, to deduce the power irradiated in the 6-13.6eV energy range \citep{marino11}. Based on the GALEX images, about 5\% of the flux is within $\sim$500\,pc from the nucleus of IC5063. This leads to an average observed, attenuated ultraviolet flux f$_{\rm UV}$ of order 1 Draine at that distance. We thus assigned the intrinsic, unattenuated f$_{\rm UV}$ values of  1, 5, or 10 Draine to allow for extinction, noting that this parameter mainly plays a role for the CO SLED for $n_{\rm H}\lesssim10^3\,{\rm cm}^{-3}$ \citep{bisbas19}. For higher densities, CRs, which exist in the ISM and are amplified in jets, can play a considerable role instead.
We allowed for CR ionization rates ($\zeta_{\rm CR}$) in the range $10^{-17}\le\zeta_{\rm CR}\le10^{-12}$\,s$^{-1}$ (with a logarithmic step of 0.1 dex). This range covers $\zeta_{\rm CR}$ values as low as those observed in the solar neighbourhood \citep{cumm15} and as high as those estimated in extreme extragalactic objects such as the type-1 Seyfert Mrk\,231 \citep{gonz18}. Typical $\zeta_{\rm CR}$ values are $10^{-16}$\,s$^{-1}$ for Milky Way regions far from the Galactic center \citep{vandishoeck_black86,mccall03, dalgarno06,neufeld10,indriolo15} and $10^{-14}$\,s$^{-1}$ near the Galactic center \citep{oka05, goto08}. To encapsulate other mechanisms of heat deposition, including shocks, we inserted a mechanical heating function with free but constant rate $\Gamma_{mech}$ \citep{loenen08}. To calculate the $\Gamma_{mech}$ range, we used the observed molecular outflow kinetic luminosity $1/2 M_{outflow} V_{outflow}^3 / d$ (as in energy-conserving starburst-driven outflows) \citep{heckman90}, distributing it in a homogeneous sphere with radius equal to the traveled distance $d$. For the outflow properties, we adopted a mass $M_{outflow}$=2$\times10^{6}$\msun\ and a velocity $V_{outflow}$=300\kms\  \citep{dasyra16,oosterloo17}. For $d$, we assumed it to be in the range 60-500\,pc. The latter value, 500\,pc, is similar to the distance between the north-west radio lobe and the nucleus. It is meaningful if the outflow started spreading from the nucleus. Alternatively, the outflow could have formed locally from the collision of the jet with a cloud $\sim$500\,pc away from the nucleus. For this scenario, we use $d$=60\,pc, which is the extent of the jet-cloud interaction region as deduced from the spatial offset between the \cooz\ peak and the \coft\ peak in the corresponding maps. This selection makes sense if 
CO(4-3) traces the densest sites of ongoing jet-cloud interaction, while CO(1-0) traces sites closer to the nucleus, containing clouds rarefied by previous interactions with the jet.
These assumptions lead to a $\Gamma_{m}$ range of  10$^{-24}$-10$^{-20}$ erg s$^{-1}$ cm$^{-3}$, which we spanned with a logarithmic step of 0.1 dex. To distinguish between the jet-related heating mechanisms, we created two additional grids: one implementing only mechanical heating ($\Gamma_{m}$=variable,  $\zeta_{\rm CR}$= $10^{-16}$\,s$^{-1}$,
as in our Galaxy), and another implementing only CR enhancement ($\zeta_{\rm CR}$=variable,  $\Gamma_{m}$=0.) \\

\begin{figure*}
	\begin{center}
		\includegraphics[width=4.25cm]{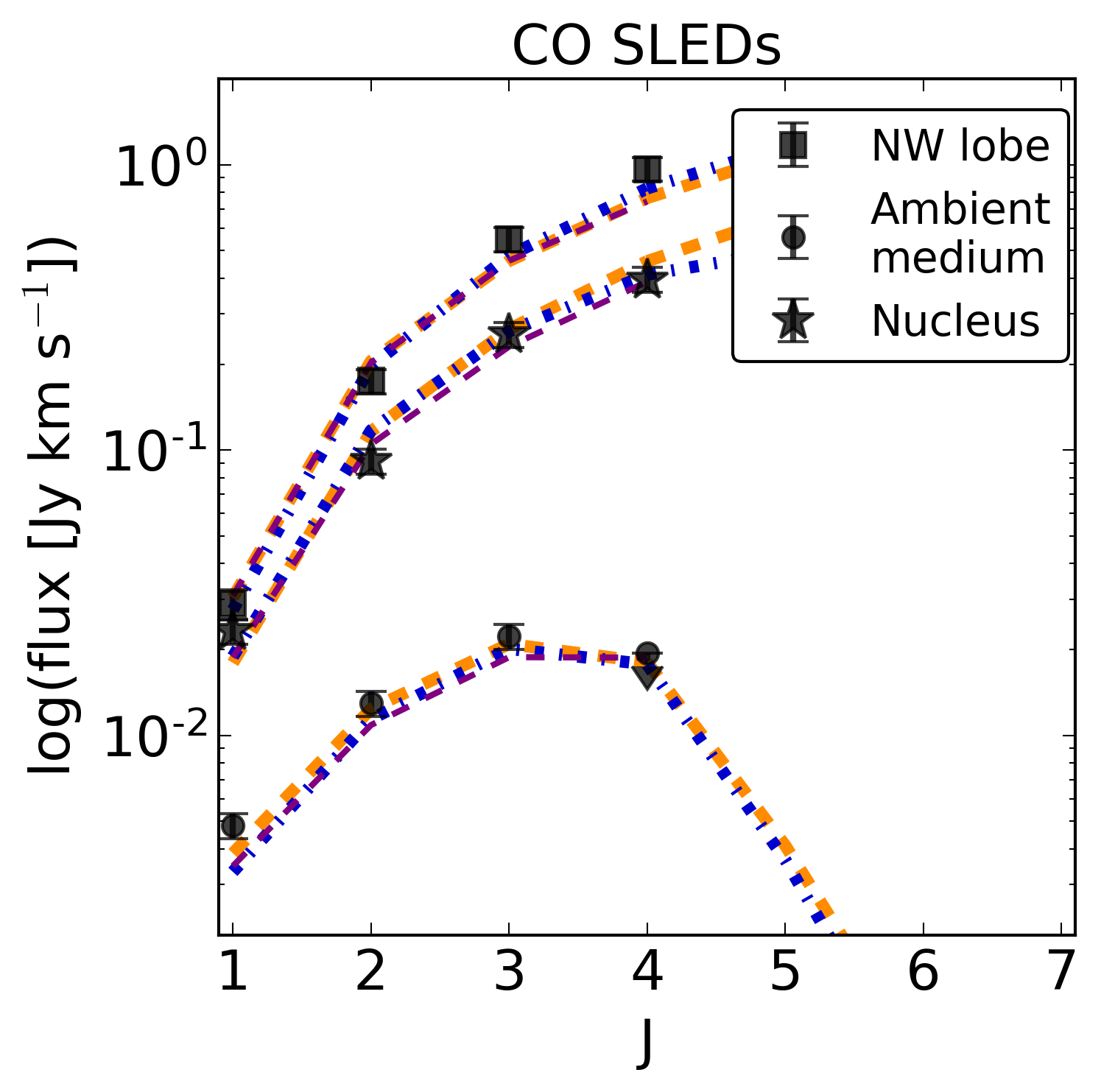}	
		\includegraphics[width=4.0cm]{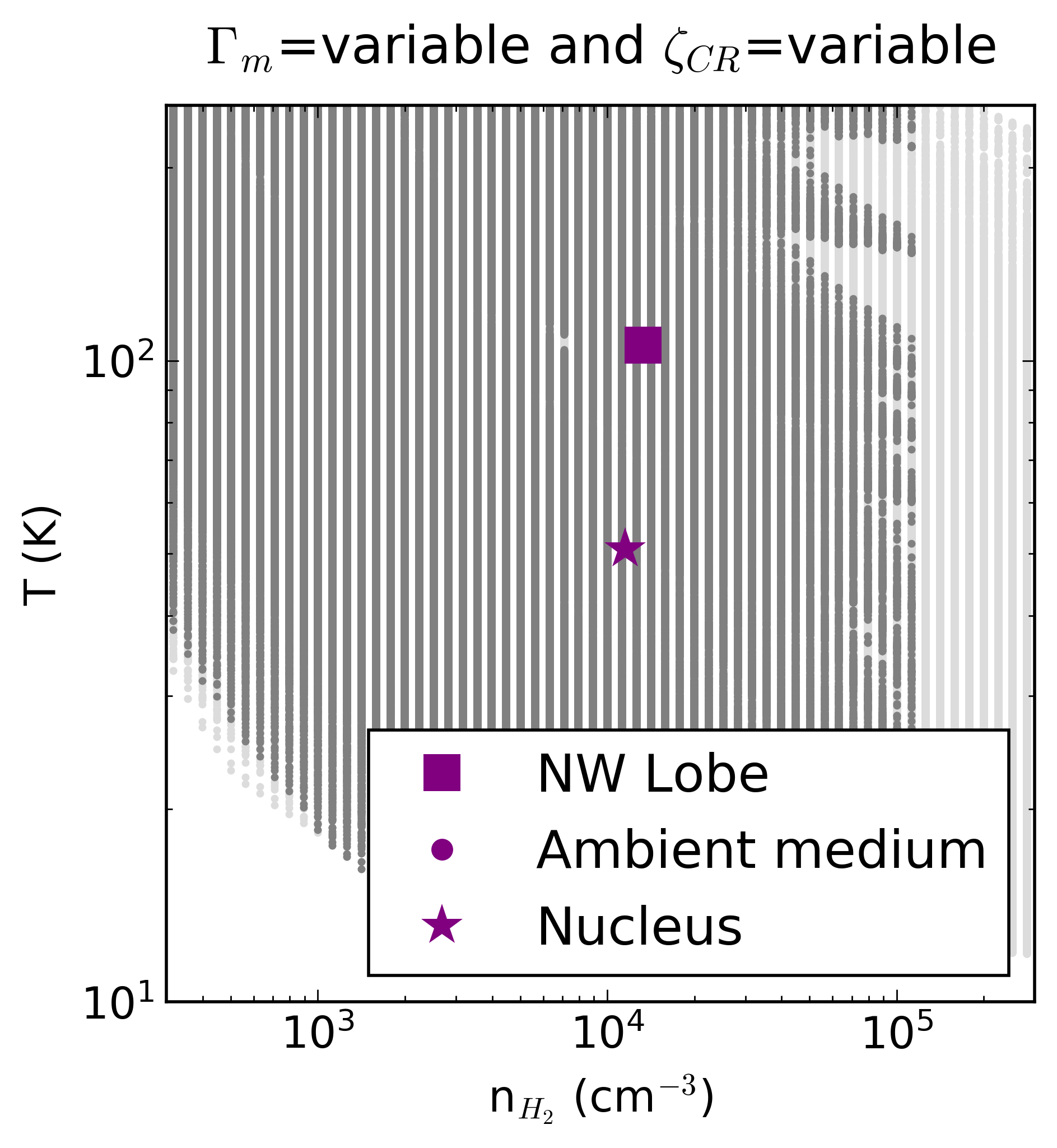}
		\includegraphics[width=4.0cm]{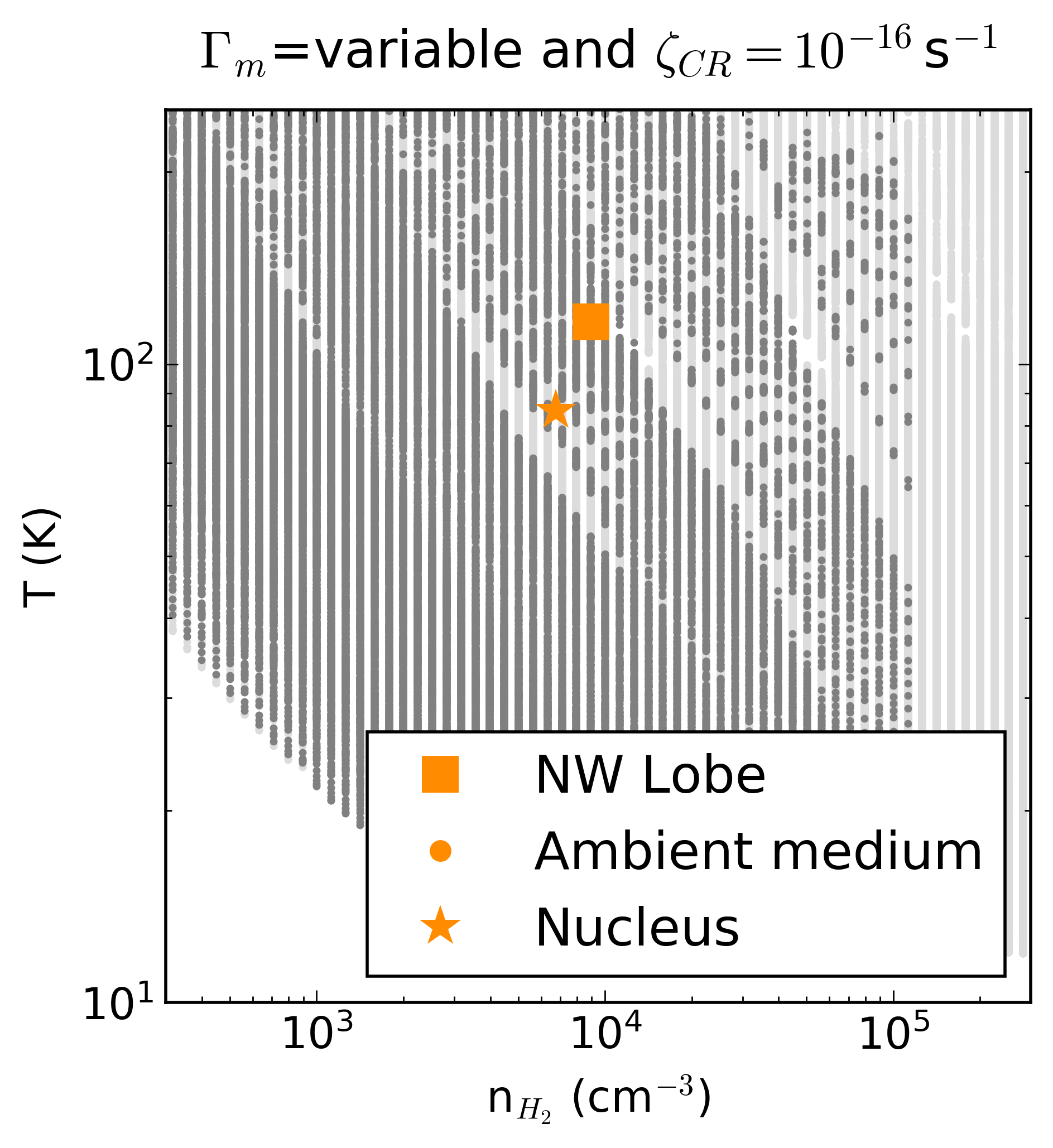}
		\includegraphics[width=4.0cm]{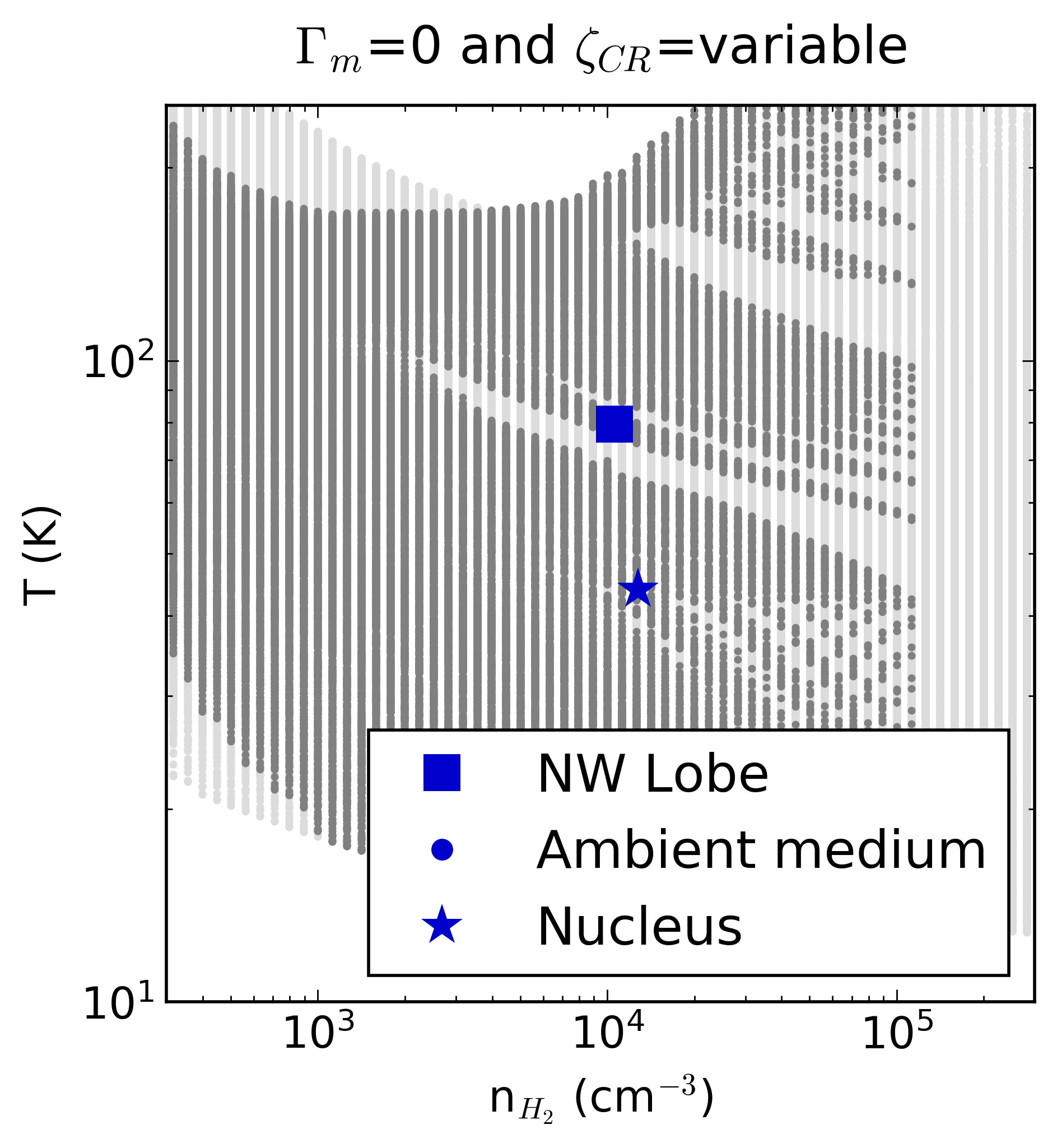}\\
		\includegraphics[width=4.0cm]{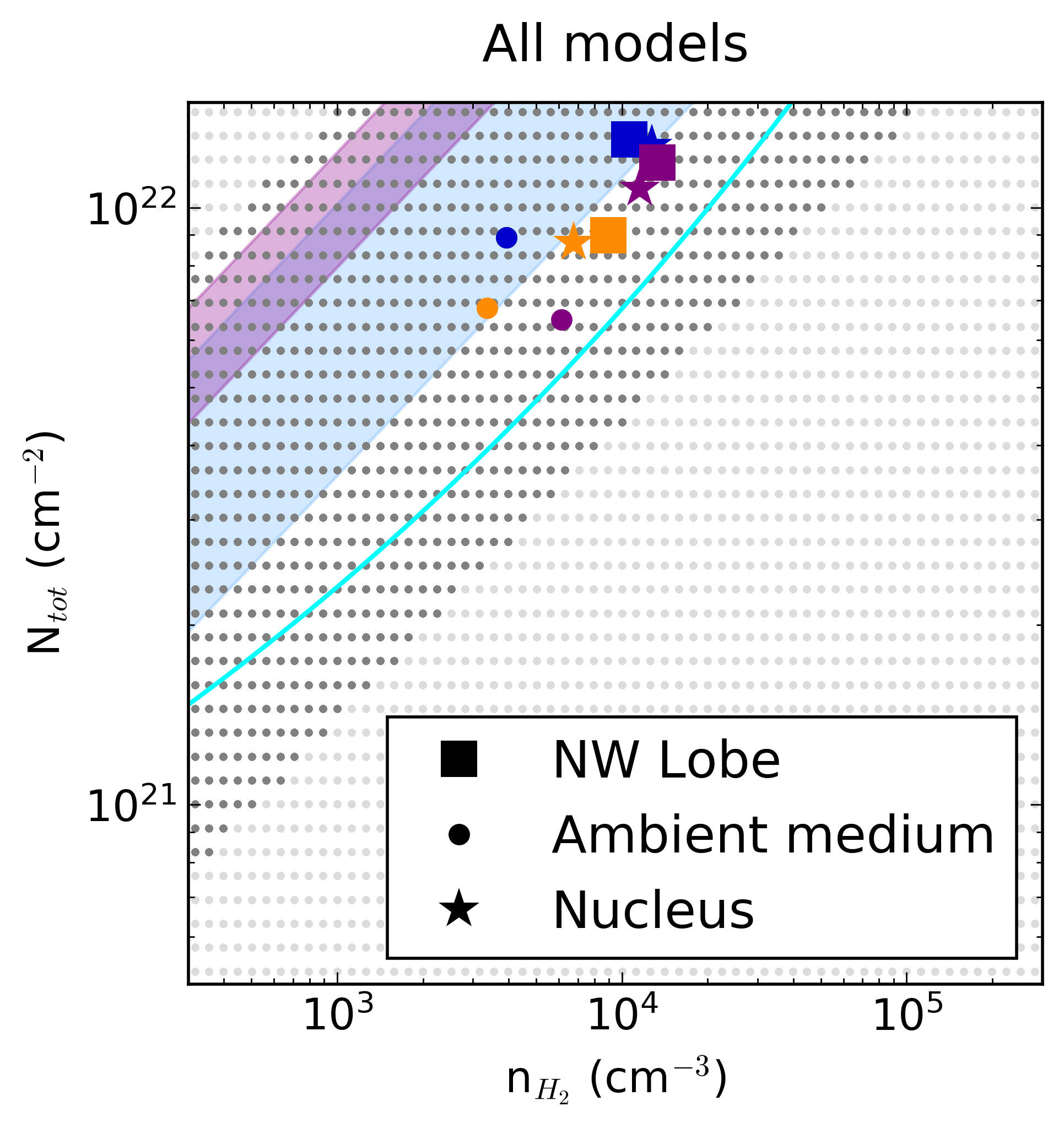}	\includegraphics[width=4.0cm]{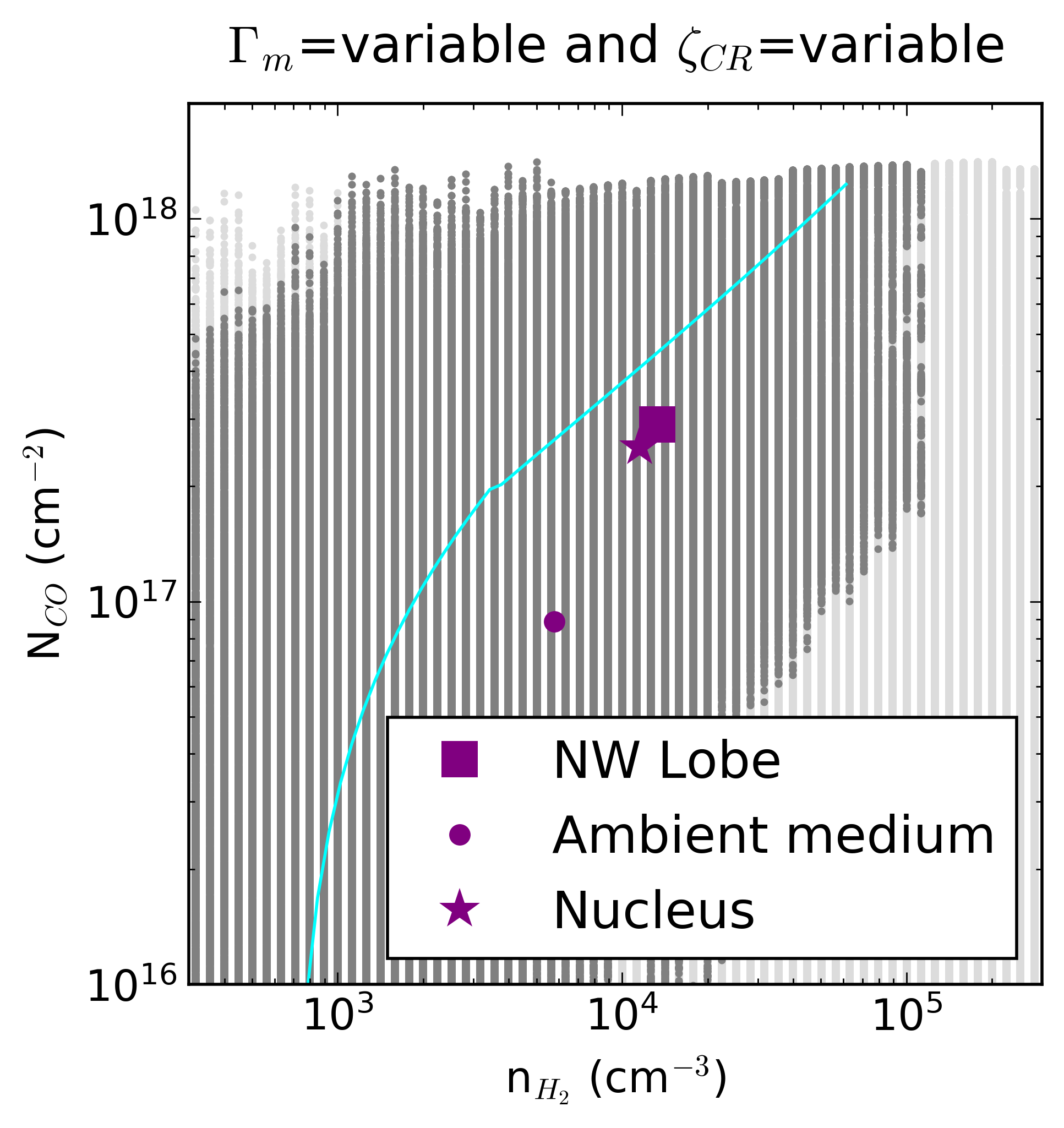}
		\includegraphics[width=4.0cm]{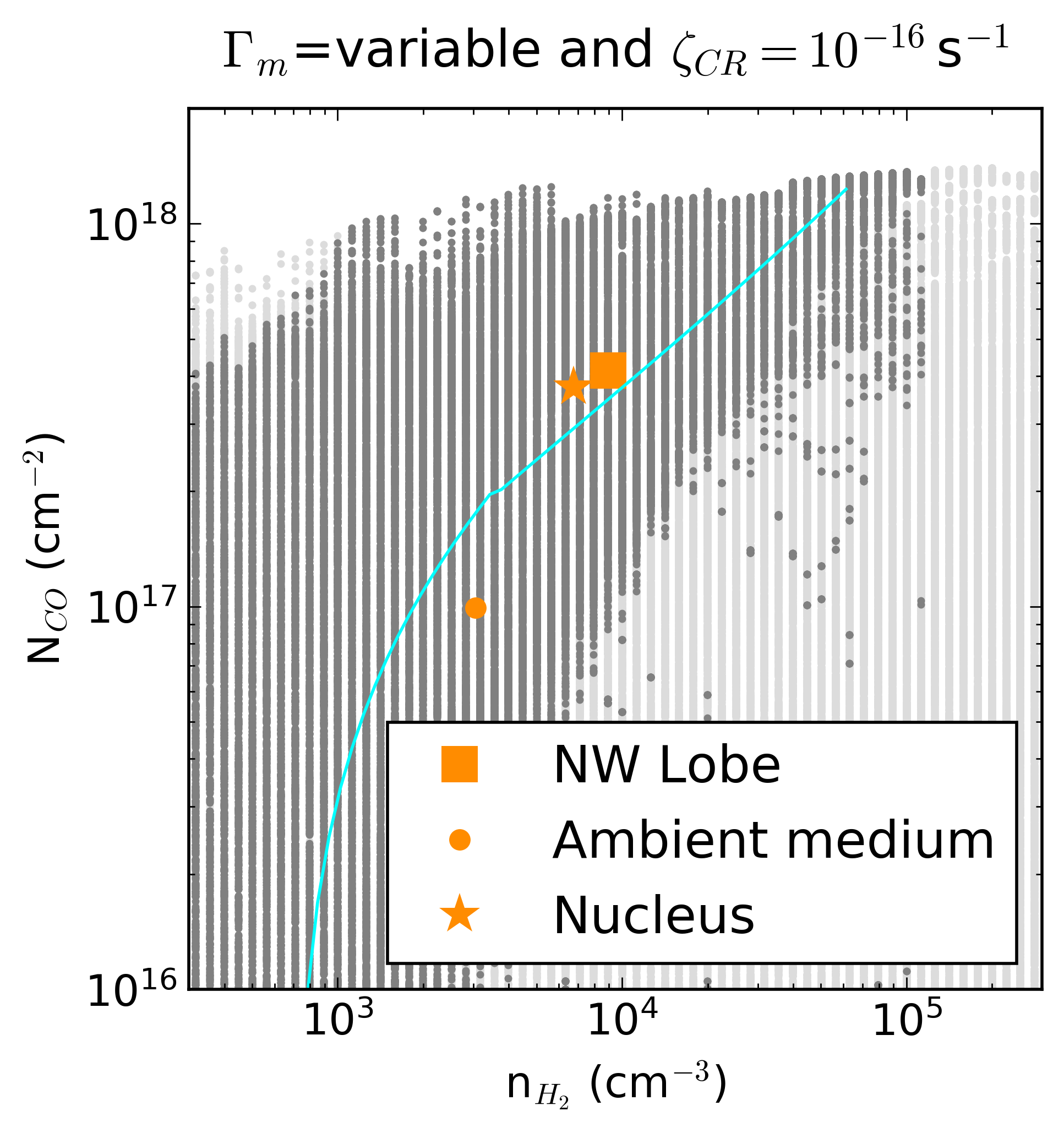}
		\includegraphics[width=4.0cm]{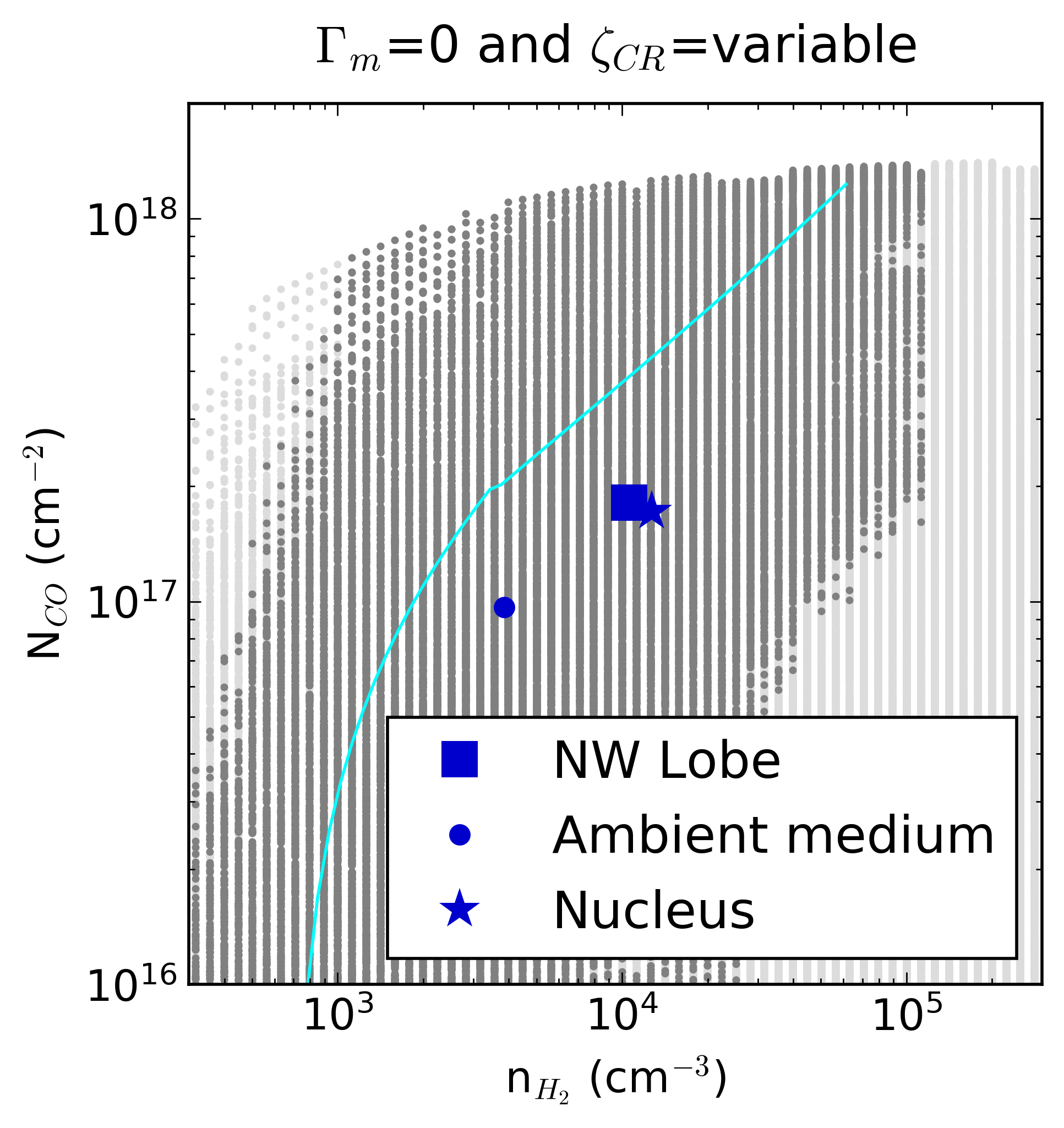}\\

    \end{center}
    \vskip -18pt
		\caption{ SLED fitting results for characteristic regions {\it Top left}: CO SLEDs. {\it Bottom left}: Corresponding \Ntot\ and \nhtwo\ solutions plotted over the input grid \Ntot -\nhtwo\ plane. The cyan curve corresponds to the $A_V$-\nh\ relation for $N_{\rm H}$/\Nhtwo = 2 and typical \Nhtwo/\Nco\ abundances \citep{smith14}. Shaded areas indicate the areas occupied by virialized clouds (see equation~\ref{eq:plane}): the blue area is for \Tkin=20\,K and 0$<$$v_{turb}$$<$1 \kms , the purple area is for \Tkin=100\,K and 0$<$$v_{turb}$$<$1 \kms . Orange, blue, and purple points correspond to  mechanical, CR, and combined heating model results, respectively. {\it Other top panels:} \Tkin -\nhtwo\ plane projection for all models. {\it Other bottom panels:} \Nco\ -\nhtwo\ plane projection for all models. The initial grids and the grids trimmed for dynamical considerations are shown in light and dark grey, respectively.
		} 
 
		\label{fig:grid}
\end{figure*}

We set the input parameters that relate to the spatial distribution and intrinsic properties of the clouds as follows. To ensure that the modeled CO lines met the observed $\sim$1\kms\ linewidth, the gas internal microturbulent velocity was set to 0.6\kms\  \citep{vandertak07}. The total H-nucleus number density, \nh, extended from 10$^2$ to 10$^6$ cm$^{-3}$ with a step of 0.05 dex. The visual extinction ${\rm A}_{\rm V}$, which relates as $6.3\times10^{-23}$ $\rm {N_H}$ $\,{\rm mag}\,{\rm cm}^2$ to the H-nucleus column density $\rm {N_H}$ \citep{weingartner01,roellig07}, had a minimum value of $10^{-3}\,{\rm mag}$, a maximum value of 10\,mag, and a step of 0.02 dex. 
Cloud dynamics further constrain the viable parameter range combinations of any grid. To select stable clouds, \citet{bisbas19} used an $A_{\rm V}$-\nh\ relation that was established by hydrodynamical simulations of molecular clouds at parsec to kiloparsec scales \citep{glover10, vanloo13, safranekshrader17, seifried17}. However, the dynamical (equilibrium or non-equilibrium) conditions of jet-impacted clouds are unknown. Only grid limits ought to be set from dynamical constraints. For this purpose, we provide an analytic expression for the stability of a molecular cloud as a function of three basic parameters of any RT code: the cloud column density, volume density, and kinetic temperature. For \htwo\ molecules, the expression is 
\begin{equation}
N_{\rm H_2}=\frac{1}{m_{\rm H_2}} \sqrt{\frac{20}{9\pi} \frac{n_{\rm H_2}}{G} (5kT_{kin}+3m_{\rm H_2} v_{turb}^2 )},
\label{eq:plane}
\end{equation}
where $G$ is the gravitational constant, $v_{turb}$ is the one-dimensional turbulent velocity of a cloud, and m$_{\rm H_2}$ is the mass of an \htwo\ molecule. The existence of this
 three-dimensional plane can be derived in a Jeans-like computation, starting from the assumption that a cloud is a virialized sphere of uniform density, supported against gravity by thermal and turbulent motions. From the virial theorem, the cloud virial mass is \hbox{$5R$\,$(5kT+3 m_{\rm H_2} v_{turb}^2)/(3Gm_{\rm H_2})$}. 
Equation~\ref{eq:plane} is obtained by equating 
 the virial mass with \hbox{$\pi R^2 \Nhtwo m_{\rm H_2} $}, where the cloud radius $R$ is $3\Nhtwo/4\nhtwo$, and solving for \Nhtwo . In this computation, we assumed that the cloud is fully molecular, that its mass is dominated by \htwo, and that its heat capacity ratio is 7/5, appropriate for linear diatomic molecules.  
 Limiting \Tkin\ and $v_{turb}$ reveals the viable parameter range in the column density - volume density plane. One limit is at \Tkin = 3\,K (the CMB temperature) and $v_{turb}$=0\kms . The other limit is at \Tkin = 200\,K, i.e., the maximum temperature indicated by an analytic solution for the \coft /\coto\ ratio in optically thin conditions and local thermodynamical equilibrium \citep{dasyra16}, and $v_{turb}$=1\kms.  The viable subgrids (Fig.\ref{fig:grid}), 
 thus, contain out-of-dynamical-equilibrium solutions, such as collapsing cores and expanding cloud shells, while encompassing the $A_{\rm V}$-\nh\ relation. \\

\begin{figure*}
	\begin{center}
		\includegraphics[width=18.4cm]{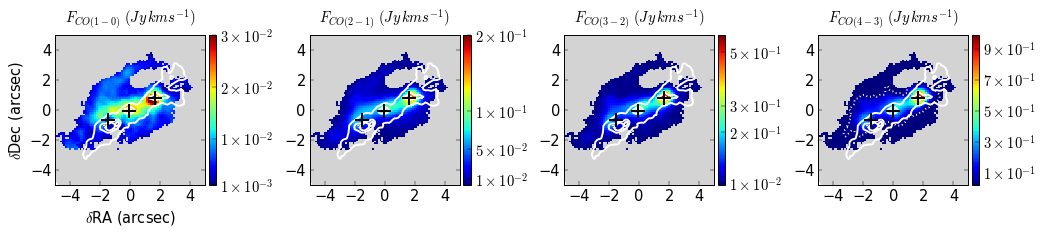}\\ 
		\includegraphics[width=4.4cm]{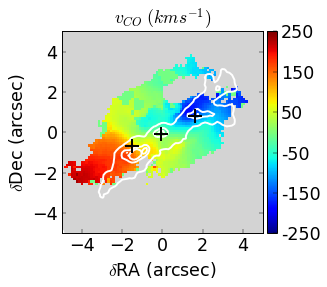} 
		\includegraphics[width=4.3cm]{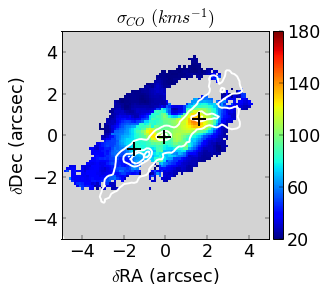}
		\includegraphics[width=4.3cm]{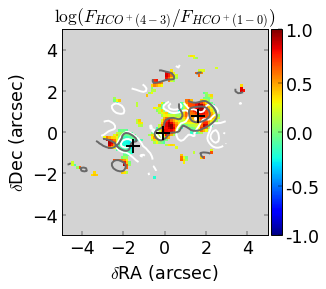}\\
	\end{center}
	\vskip -18pt
		\caption{ Observational maps used for the spatially-resolved SLED fitting. {\it Upper panels:} line fluxes. Only pixels with S/N ratio $>$3 in CO (1$-$0), (2$-$1) and (3$-$2) are shown. For \coft , the dotted line traces the S/N=3 contour.  Upper limits were used for pixels outside this contour. 
		{\it Lower panels}: Velocity (left) and velocity dispersion (middle) averaged for all CO lines, with contours marking the jet trail as seen in the optical.
		 HCO$^+$ (4$-$3)/(1$-$0) ratio (right), with the 1$-$0 emission shown with white contours of S/N=5,\,10, and the 4$-$3 emission shown with
         black contours of S/N=5,\,20.
			} 
		\label{fig:line_input}
\end{figure*}

To select the optimal model, we performed an error-weighed $\chi ^2$ minimization between observed and modeled CO line fluxes in logarithmic space \citep{viti14}, which yields better results than a minimization in linear space for fluxes of different orders of magnitude. Prior to the minimization, we divided the observed fluxes by a beam filling factor \bff, which parameterizes the fraction of a telescope beam occupied by clouds, which converts observed beam-averaged brightness temperatures to intrinsic brightness temperatures, and which is common for all CO lines. We allowed \bff\ to vary between $1\times10^{-3}$ and $0.1$, with a step of 0.5 dex. We ran the minimization algorithm for all CO spectral lines simultaneously requesting that the solution also satisfies the HCO$^+$ (4$-$3)/(1$-$0) flux ratio, when available. Owing to the use of a line ratio, the HCO$^+$ \bff\ is irrelevant for our analysis. The fitting was performed for all pixels with signal-to-noise (S/N) ratio exceeding 3 in CO (1$-$0), (2$-$1), and (3$-$2). A common mask was created for these pixels. In \coft , the detection is confined close to the jet trail (Fig.~\ref{fig:line_input}); to enable the study of quiescent regions, we used flux upper limits for all pixels between the common mask and the \coft\ S/N=3 contours (see Fig.~\ref{fig:line_input}). The best-fit solutions for characteristic regions are shown in Fig.~\ref{fig:grid}. \\

\begin{figure*}[!ht]
	\begin{center}
		\includegraphics[width=18cm]{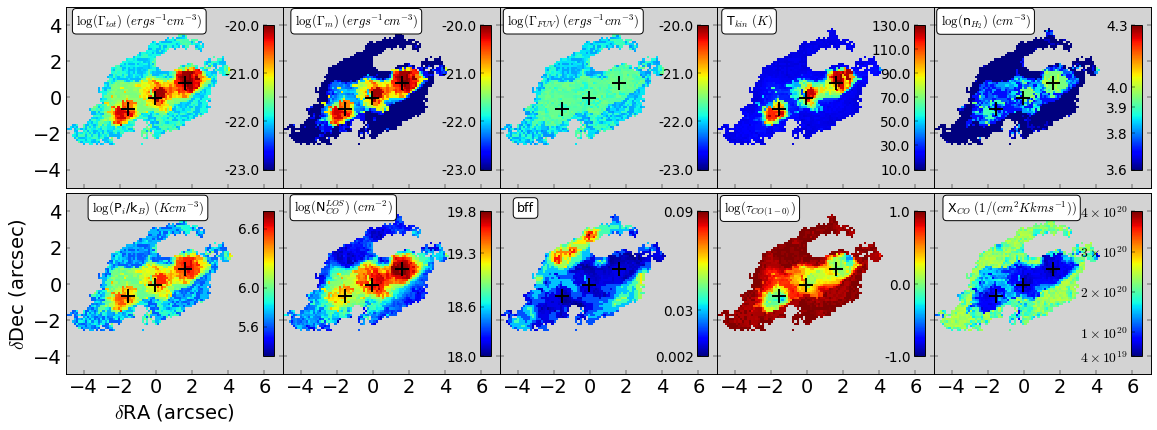}\\
		\vspace{-0.3cm}
	\end{center}
		\caption{ Spatially-resolved SLED fitting results for the model with $\Gamma_{m}$=variable and $\zeta_{\rm CR}$=$10^{-16}$\,s$^{-1}$. The quantities shown are the heating rate (total, mechanical, and FUV), the molecular gas kinetic temperature and volume density (upper row), and the cloud internal pressure, the CO column density along the line of sight, the CO beam filling factor, the CO(1-0) optical depth, and X${_{CO}}$ (lower row).
		}
		\label{fig:M_panel}
\end{figure*}

\begin{figure*}[!ht]
	\begin{center}
		\includegraphics[width=18cm]{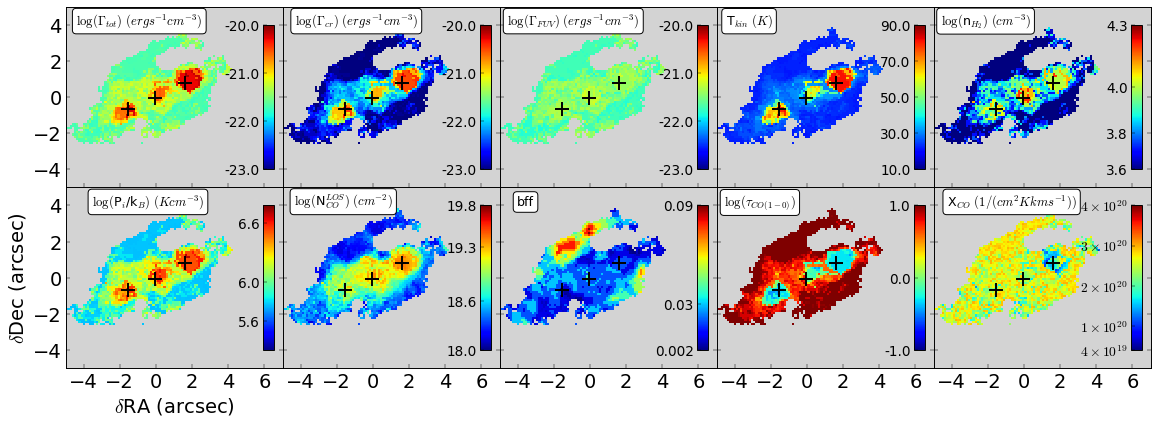}\\
		\vspace{-0.3cm}
	\end{center}
		\caption{ Spatially-resolved SLED fitting results for the model with $\zeta_{\rm CR}$=variable and $\Gamma_{m}$=0. The quantities shown are as in Fig.~\ref{fig:M_panel}.
		}
		\label{fig:CR_panel}
\end{figure*}

\begin{figure*}[!t]
	\begin{center}
		\includegraphics[height=4cm]{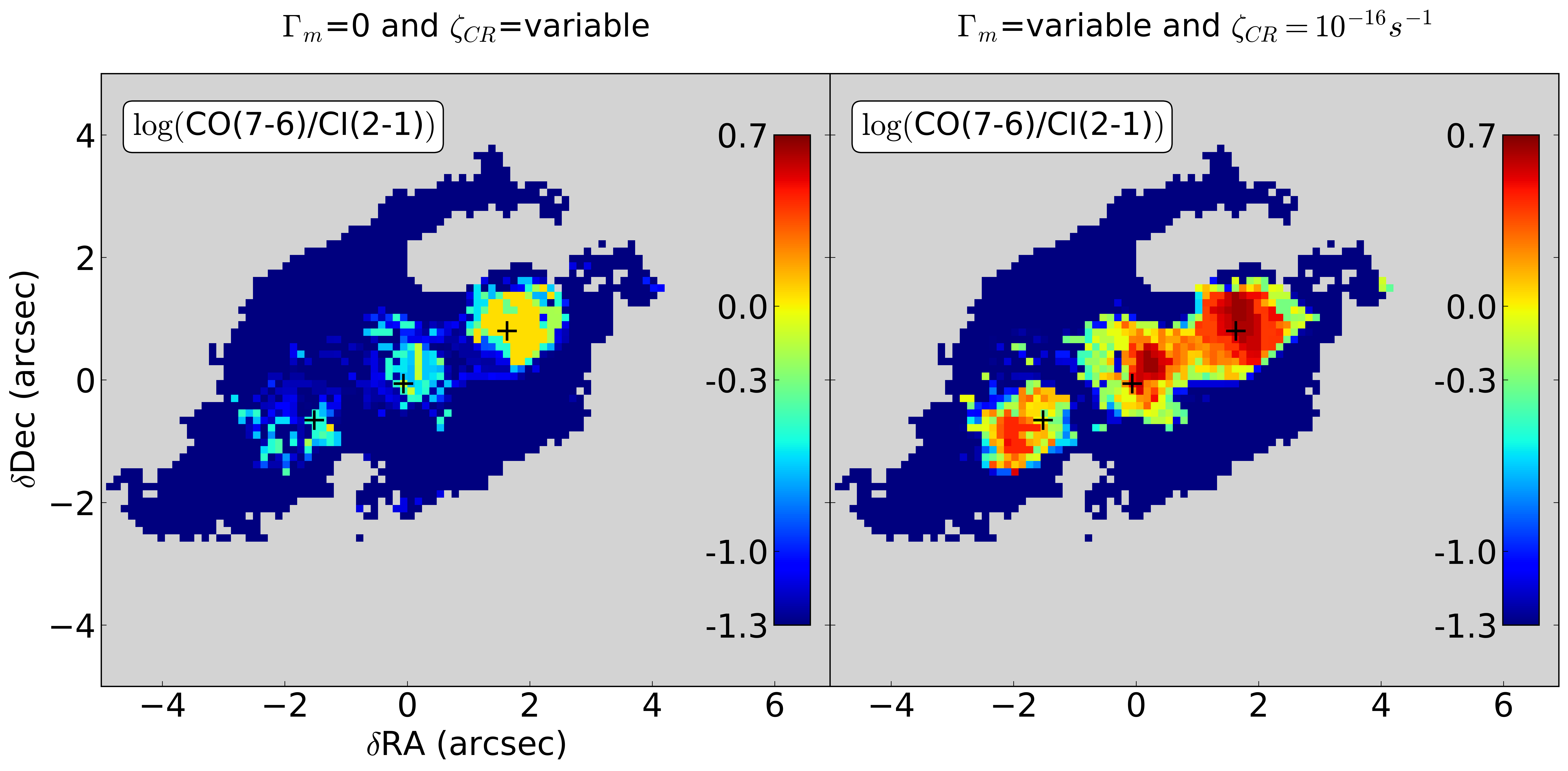}
	\end{center}
		\caption{Predictions of the CO(7-6)/CI(2-1) flux ratio distinguishing mechanical from CR heating.
		}
		\label{fig:line_predictions}
\end{figure*}

For comparison with the results of the  $\Gamma_{m}$=variable model (Fig.~\ref{fig:M_panel}), the results of the $\zeta_{CR}$=variable model are shown in Fig.~\ref{fig:CR_panel}. For both models, the results are nearly identical for the pressure and similar for most other parameters, with CRs leading to emission from somewhat cooler and denser ISM. Ambient clouds in the spiral arm are the closest to their theoretical virial state, while jet-impacted clouds at the radio lobes are the furthest away from their virial state. The change in the cloud surface is reflected in the $X_{CO}$ maps, showing whether we see more or less dissipated cloud layers.  For CRs, cold, dense gas layers and warm, shallow gas layers are likely to both contribute to the emission, leaving no impact to the overall $X_{CO}$. In all models, the typical values of the beam filling factor were between 0.01-0.1, with most clouds being (reasonably) located where spiral arms meet the central disk. Predictions of the CO(7-6)/CI(2-1) line ratio that can serve as diagnostic of the most appropriate model are shown in Fig.~\ref{fig:line_predictions}.

\subsection{Optical and NIR data}
\label{sec:opt_IR_ data}

To determine the properties of the diffuse gas layers surrounding the CO-emitting dense molecular clouds, we analyzed Very Large Telescope MUSE and SINFONI data cubes in optical and NIR wavelengths that are publicly available in the European Southern Observatory archive. The MUSE data, taken for the program 60.A-9339, were processed by the ESO MUSE pipeline version 1.6.1. The cube that we obtained is nearly identical to that produced when manually running the ESO Reflex routines \citep{venturi21}. We fitted the profile of each line in the MUSE data with two Gaussian functions, one for the disk and one for the outflow, as required for the deblending of blended lines. Therefore, four Gaussians were used for the fitting and proper flux measurement of \sii\ 6716\AA\ and 6731\AA, which overlap in regions with outflows. An extinction correction was also applied for lines far in wavelength, for example for \nii \ at 5755\AA\ and for the \nii \ doublet at 6548\AA\ and 6584\AA . The correction was derived from the Balmer ratio, assuming an intrinsic H$\alpha$/H$\beta$ decrement of 3.1 \citep{osterbrock89}, a foreground extinction, and a Milky Way extinction curve \citep{cardelli89}. Six Gaussians were, in total, used for the deblending of the H$\alpha$ and above-mentioned \nii\ doublet complex, which was fitted simultaneously with (another two Gaussians for) H$\beta$. Similar kinematics were assumed for \nii \ 6548\AA\,and 6584\AA\,on one hand,\,and H$\alpha$ and H$\beta$ on the other. Both the disk and the outflow of these  same-species lines were assumed to have the same central velocity and velocity dispersion. Prior to fitting all gas emission lines, we created a template of the typical spectrum emitted by all stellar populations in the center of IC5063 from gas-free regions of the cube (from 13-32\arcsec\ away from the jet trail, parallel to it). The template was scaled and removed from each pixel's to ensure that stellar absorption does not alter the emission line flux measurements. \\

In the SINFONI data, which were acquired for the programs
075.B-0348A and 089.B-0971A and reduced according to the recipes in their initial publication \citep{dasyra15}, the smaller field of view of the instrument did not allow for a similar removal of the stellar absorption. Instead, we removed the CO(7-4) absorption shortward of  \feii\ 1.644 \micron\  using a rescaled CO(6-3) bandhead profile. The use of CO(2-0) instead of CO(6-3) yielded indistinguishable results. Similarly, the \feii\ 1.533 \micron\ line was contaminated by weak Brackett 18-4 emission. To remove it, we subtracted the Brackett 13-4 line profile from the data cube after scaling its peak to the Brackett 18-4 peak.  \\

\begin{figure*}[!ht]
	\begin{center}
		\includegraphics[width=0.28\textwidth]{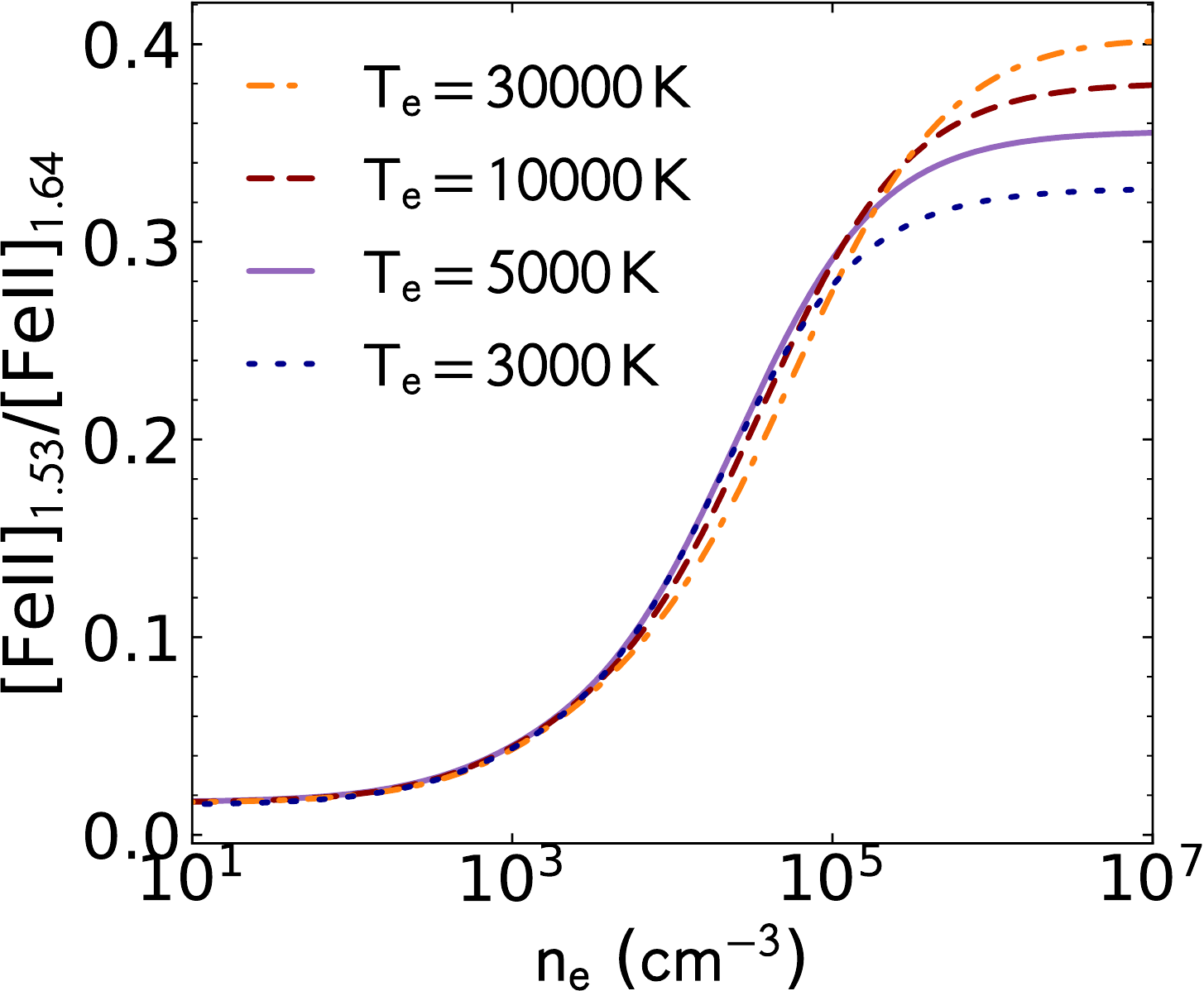}~		\includegraphics[width=0.28\textwidth]{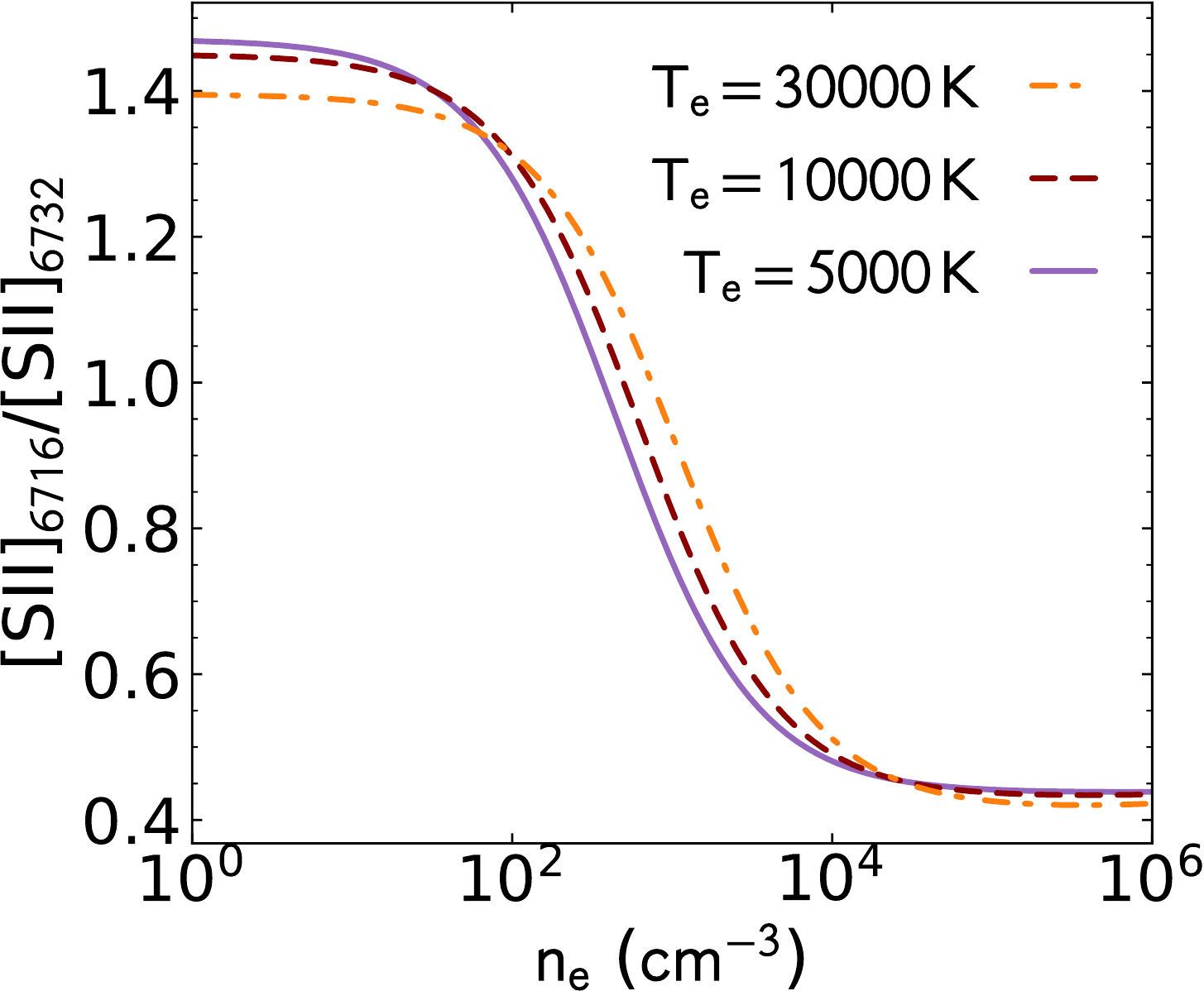}~
		\includegraphics[width=0.29\textwidth]{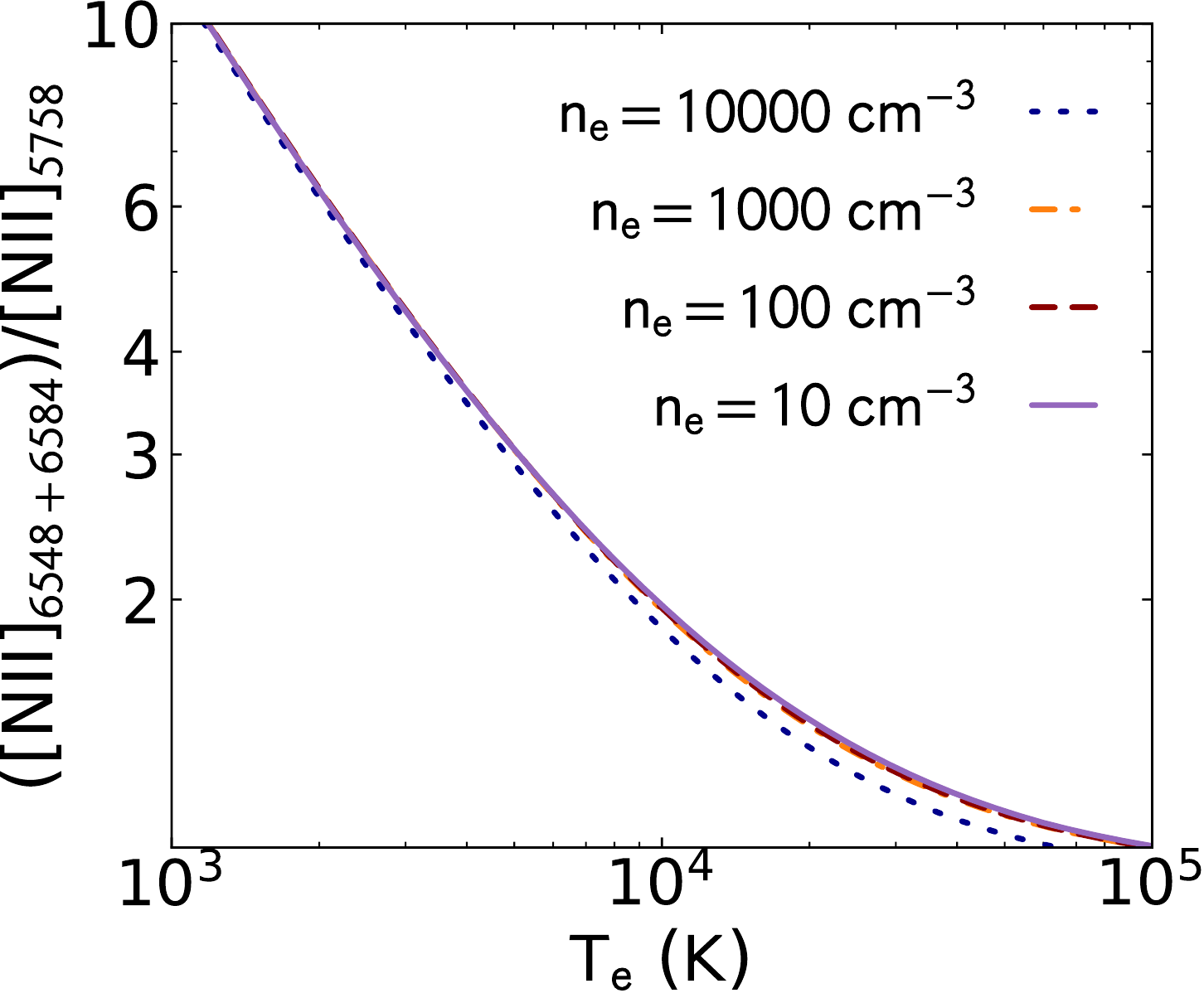}
     	\vspace{-0.3cm}
	\end{center}\
		\caption{ \feii\ and \sii\ flux ratios used as density diagnostics {\it (left and center)} and \nii\ flux ratio used as temperature diagnostic {\it(right)} as obtained by PyNeb \citep{luridiana15}.
		}
		\label{fig:line_diagnostics}
\end{figure*}

\begin{figure*}[!t]
	\begin{center}
		\includegraphics[height=5.7cm]{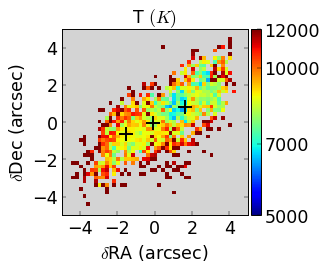}
		\includegraphics[height=5.7cm]{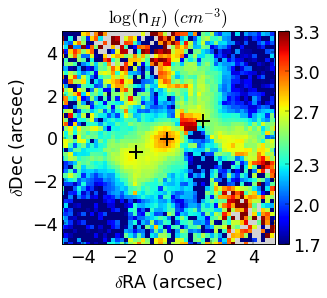} 		\\
	\end{center}
		\caption{Optical gas properties derived from the VLT MUSE data.  {\it Left}: Temperature map based on the \nii\ flux ratio. {\it Right}: Density map based on the \sii\ flux ratio.
		}
		\label{fig:optical_pressure}
\end{figure*}

In the optical regime, the flux ratio \sii\ 6716\AA /\sii\ 6731\AA\ is a classical density probe, involving transitions of comparable energy that result in the same electronic quantum level  \citep{osterbrock89}. The code PyNeb \citep{luridiana15} was used for the computation of this ratio as a function of density with updated atomic coefficients (Fig.~\ref{fig:line_diagnostics}), providing results similar to those of other recent codes
\citep{kewley19}. The deblended \sii\  line ratio map from the MUSE data was used to produce the density map shown in Fig.~\ref{fig:optical_pressure}, under the assumption that the optically-emitting ionised gas is at 10.000K. The results are, nonetheless, little dependent on this temperature. Temperatures were measured using the auroral \nii \ line at 5755\AA \ and the \nii \ doublet at 6548\AA, 6584\AA, since the brighter oxygen lines cannot be used instead (as \oiii \ at 4363\AA \ is out of the MUSE spectral range). Again, T was derived using PyNeb, from the \nii 6548\AA +\nii 6584\AA /\nii 5755\AA \ ratio, and assuming $n_e = 100\, \rm{cm^{-3}}$. Expected variations with density are negligible as shown in Fig.~\ref{fig:line_diagnostics}. The results for the temperature are also shown in Fig.~\ref{fig:optical_pressure}.\\

\begin{figure*}[!t]
	\begin{center}
		\includegraphics[height=5.7cm]{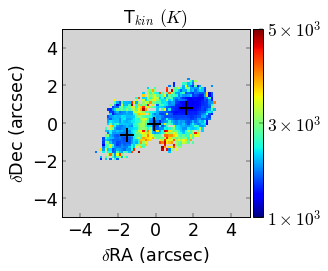}
		\includegraphics[height=5.7cm]{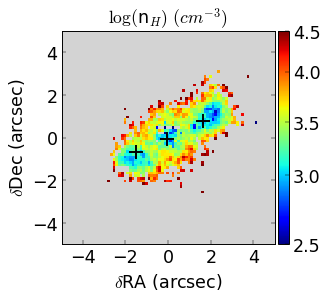} \\

	\end{center}

		\caption{NIR gas properties derived from the VLT SINFONI data. {\it Left}: Temperature based on the \htwo\ (1-0) S(1) and S(3) line fluxes. The highest \Tkin\ values are close to regions of high turbulence  perpendicular to the jet, associated with it \cite{dasyra15,venturi21}.{\it Right}: Density based on the \feii\ 1.533 and 1.644 \micron\ line fluxes. }
		\label{fig:nir_pressure}
\end{figure*}

In the NIR, the \htwo\ rovibrational lines (1-0) S(1) at 2.1218\micron\ and (1-0) S(3) at 1.9576 \micron\ were used for the analytic determination of \Tex\ of the warmest molecular gas layers, under the assumption that this gas is optically thin and in local thermodynamic equilibrium \citep{dasyra15}. \Tkin\ is then equal to \Tex , which is found by comparing the measured to the theoretical luminosity ratio of the two lines. The theoretical luminosity is the product of the Boltzmann-distribution-related number of photons emitted during a transition times the photon energy times the transition Einstein coefficient. The \Tex\ map for the \htwo -emitting gas indicates a temperature range of 1500-4000\,K (Fig.~\ref{fig:nir_pressure}), with fluorescence adding uncertainty to \Tex\ because of its potential contribution to the emission. Still, the average \Tex\ lies in the lower part of this range in most regions \citep{dasyra15}. Higher values are mainly seen in regions with high fractions of outflowing gas, e.g., in the biconical \htwo\ outflow \citep{dasyra15}. A density probe in the NIR that is effective for gas of several thousand Kelvin is the flux ratio \feii\ 1.533\micron /\feii\ 1.644\micron. These lines probe transitions from the same upper electronic level to lower electronic levels of comparable energy, so that the main factor that determines the number of ions that de-excite radiatively relates to the critical density. Their critical densities differ enough that their ratio can be used as a density measure in the range 10$^2$-10$^6$ cm$^{-3}$ \citep{pradhan_zhang93}. We obtained an updated analytic function that connects the \feii\ 1.533\micron / \feii\ 1.644\micron\ flux ratio with the ISM density by running PyNeb (Fig.~\ref{fig:line_diagnostics}). Our calculations indicated ionised gas densities in the range 300-30000 cm$^{-3}$ (Fig.~\ref{fig:nir_pressure}), for a typical temperature of 5000\,K. This temperature is well chosen for two reasons. Firstly, the \feii\ ratio was limited to a maximum of 0.3, not requiring higher temperatures to be produced (Fig.~\ref{fig:line_diagnostics}). Secondly, the \feii\ emission is cospatial with the few-thousand-Kelvin \htwo\ gas in IC5063 \cite{dasyra15}. The derived density appears to be lowest in the jet-impacted regions and to increase with distance from the jet trail (Fig.~\ref{fig:nir_pressure}). This overdensity, which needs to be confirmed with higher S/N ratio infrared data, could be a nice tracer of the border of the cocoon, which shows a mild CO outflow \cite{morganti17}.  \\

\acknowledgments
{\noindent \bf Aknowledgements:} KMD, GFP, and TGB acknowledge financial support by the Hellenic Foundation for Research and Innovation (HFRI), under the first call for the creation of research groups by postdoctoral researchers that was launched by the General Secretariat For Research and Technology (project number 1882). TGB further acknowledges support from Deutsche Forschungsgemeinschaft (DFG) grant No. 424563772. JAFO acknowledges financial support by the Agenzia Spaziale Italiana (ASI) under the research contract 2018-31-HH.0. GFP is supported for this research by the International Max-Planck Research School (IMPRS) for Astronomy and Astrophysics at the University of Bonn and Cologne.
This paper makes use of the ALMA data 2012.1.00435.S, 2015.1.00420.S, 2015.1.00467.S,  2016.1.01279.S. ALMA is a partnership of ESO (representing its member states), NSF (USA) and NINS (Japan), together with NRC (Canada) and NSC and ASIAA (Taiwan) and KASI (Republic of Korea), in cooperation with the Republic of Chile. The Joint ALMA Observatory is operated by ESO, AUI/NRAO and NAOJ.\\

\clearpage

\vskip -18pt

{}

\end{document}